# Wide field fluorescence epi-microscopy behind a scattering medium enabled by speckle correlations

Matthias Hofer,[1] Christian Soeller,[2] Sophie Brasselet,[1,*] and Jacopo Bertolotti[2,†]

[1] Aix Marseille Univ, CNRS, Centrale Marseille, Institut Fresnel, F-13013 Marseille, France
[2] University of Exeter, Stocker Road, Exeter EX4 4QL, United Kingdom
*sophie.brasselet@fresnel.fr
†j.bertolotti@exeter.ac.org

**Fluorescence microscopy is widely used in biological imaging, however scattering from tissues strongly limits its applicability to a shallow depth. In this work we adapt a methodology inspired from stellar speckle interferometry, and exploit the optical memory effect to enable fluorescence microscopy through a turbid layer. We demonstrate efficient reconstruction of micrometer-size fluorescent objects behind a scattering medium in epi-microscopy, and study the specificities of this imaging modality (magnification, field of view, resolution) as compared to traditional microscopy. Using a modified phase retrieval algorithm to reconstruct fluorescent objects from speckle images, we demonstrate robust reconstructions even in relatively low signal to noise conditions. This modality is particularly appropriate for imaging in biological media, which are known to exhibit relatively large optical memory ranges compatible with tens of micrometers size field of views, and large spectral bandwidths compatible with emission fluorescence spectra of tens of nanometers widths.**

Scattering in heterogeneous disordered media scrambles light, blurring any image to the point of uselessness. One can recover a sharp image by rejecting the scattered light [1,2] but the amount of viable signal decreases exponentially with the optical density, making this approach only viable for thin or semitransparent systems. This is particularly problematic for epi-fluorescence imaging in biological tissues, where scattering often prevents imaging beyond a shallow depth [3]. A common approach to this problem is the use of adaptive optics, where a "guide star" close to the object to be imaged is used to fully characterize the point spread function (PSF) of the imaging process due to scattering, which is then corrected for [4]. But, as adaptive optics essentially corrects for ballistic light, it is unable to retrieve an image when the scattering is strong enough to generate a fully developed speckle pattern [5]. If the scattering medium can be fully characterized beforehand the light scrambling can be inverted and a sharp image recovered [6–8], however this requires accessing both sides of the medium, which is almost never met in fluorescence microscopy experiments with biological samples.

It has been recently realized that even though a speckle pattern appears to be spatially random, it exhibits correlation properties with respect to the angle of illumination of the scattering medium. This optical memory effect has been exploited to retrieve the image of a fluorescent object behind a completely opaque scattering layer [9,10]. The object's autocorrelation was obtained scanning the angle of incidence of an excitation beam and measuring the total fluorescence that managed to pass through the scattering layer. The autocorrelation was then inverted using an iterative algorithm [11] to yield an image of the object. Subsequently it has been realized that, provided that the light is temporally coherent and spatially incoherent, the lengthy pump scanning is unnecessary and the measurement can be done in a single shot [12]. Furthermore, other forms of speckle correlation were described and exploited for imaging [13,14]. All those demonstrations have relied on high power lasers combined with narrow band detected spectral profiles to ensure high signal to noise (SNR) and temporal coherence conditions. In addition, a transmission geometry was used to ensure a better control of the object illumination homogeneity. These experimental conditions are however not compatible with epi-fluorescence microscopy, which is found in the majority of applications to biology and medical sciences.

In this work we show that micrometer-scale fluorescent objects hidden behind a scattering layer can be imaged non-invasively in a regular inverted wide-field microscope with single shot acquisition. The use of a microscope objective allows for efficient detection of the scattered light up to hundreds of micrometers away from the scattering medium and guarantees sufficient sampling on the camera. The optical diffuser used as a scattering medium in this work exhibits spectral properties which resemble those of biological samples of thicknesses below a millimeter, as frequently used in fluorescence microscopy. In this regime, we show the possibility to enhance signal to noise conditions of fluorescence imaging by increasing the measured spectral bandwidth, benefiting from the large spectral bandwidth of the medium, which is defined as the spectral range over which the detected speckle doesn't change significantly. Finally, we present a novel combination of Fienup-type algorithms that is computational efficient and stable at fairly low signal-to-noise ratios, which is critical to make this technique practical under common laboratory conditions.

**Materials and methods**
The experiments were conducted using a regular inverted microscope (Axio Observer.D1, Carl Zeiss AG) with a continuous wave diode-pumped solid state laser (532-DPSS, Viasho) as excitation light source that emits light at a wavelength of 532 nm (Fig. 1(a)). The laser beam was expanded to a diameter of 4 mm by a Galilean beam expander with a magnification of 2.5. A 565LP dichroic mirror served to separate excitation from emission light. A 10x/NA=0.25 objective (Olympus Plan N) was used to excite the fluorophores and to collect the fluorescence emission. Different fluorescent samples were used. In a first type, orange fluorescent beads (FluoSpheres™, 1.0 μm, Thermo Fisher Scientific Inc.) were deposited on a glass coverslip and imaged using a 590/10 bandpass emission filter. For controlled shapes of smaller sizes, regular micrometric scale structures were made by depositing and cross-linking quantum dots coated with an organic coating (Qdot™ 705 ITK™, Thermo Fisher Scientific Inc.), where ion-beam scanning assisted cross-linking was used to generate well-controlled patterns. QD emission was observed through a 700/10 bandpass filter, unless indicated otherwise. A 600grit ground glass diffuser (Thorlabs) served as the scattering medium. The scattered fluorescence emission was detected by a scientific CMOS camera (Zyla 4.2 PLUS sCMOS, Andor Technology Ltd.) with a dynamic range of 16 bits. Integration times were adapted to make use of the full dynamic range of the camera. In some experiments a bandpass filter was used to control the spectral bandwidth (i.e the temporal coherence) of the light reaching the camera sensor. Both the distance from the object to the diffuser and from the diffuser to the objective focal plane were controlled by translation stages (Fig. 1(b)). The set-up used in this work allowed imaging the object *in situ* before insertion of the diffuser, permitting a direct comparison of the actual object fluorescence pattern with the reconstructed object fluorescence. An example of such an object, made of fluorescent beads, is shown in Fig. 1(c). The pattern exhibits fine details in the micrometer range and a spatial complexity not unlike the distributions observed in fluorescently stained biological cells and tissues.

**Principle of the image retrieval**
A scattering medium scrambles light, but does not completely inhibit light propagation. In the strongly scattering (diffusive) regime, the amount of light transmitted through a disordered layer decreases linearly with the optical density [15], but a fraction of the initial power is available to excite a fluorescent object. If the fluorescent light has sufficient temporal coherence, which degree can be controlled by reducing the spectral bandwidth of the emitted light, the fraction that crosses the scattering layer and the bandpass filter will form a speckle pattern [5]. However, as the fluorescent light is spatially incoherent, the speckle patterns originating from different points in the sample will sum incoherently. If the sample fits within the angular optical memory effect range [6,16] of the scattering layers, all the speckle patterns from different points are the same, and the intensity pattern I we can measure is the convolution (indicated with $*$) between the object O and the (speckle) point spread function S [12]:

$$I = O * S = \int O(x) S(x - \delta x) dx. \qquad (1)$$

By computing the autocorrelation (indicated with $\otimes$) of the measured intensity pattern we can separate the contribution of the speckle from the object [9]:

$$I \otimes I = [O \otimes O] * [S \otimes S] \cong O \otimes O. \qquad (2)$$

The autocorrelation of the object can be numerically inverted using an iterative algorithm [11] to yield a high resolution image [9,12].

**Image processing**
The recorded raw images were normalized by a Gaussian low-pass filtered version of itself to flatten the intensity envelope which results from the scattering process (Fig. 1(d)). The low-pass filter has to be adapted to the object's dimensions that can be estimated by the size of the autocorrelation (Fig. 1(e)). Subsequently, the autocorrelation image was offset corrected by subtracting its minimum value. A cosine window was chosen to cut the outer parts of the autocorrelation that contain only information on the residual noise due to an incomplete averaging of the speckle (Fig. 1(e)). According to the Wiener-Khinchin theorem, the autocorrelation of a signal represents its power spectrum. Therefore, the Fourier transform of the autocorrelation gives the modulus of the object's Fourier transform,

$$|F\{O\}| = \sqrt{|F\{I \otimes I\}|}. \qquad (3)$$

Thus, the retrieval of the correct phase in the Fourier domain will lead to the reconstruction of the object which was hidden behind the scattering medium.

In order to restore the phase in the Fourier domain, we implemented a combination of Fienup's error reduction algorithm (ER) [11] and of a modified hybrid-input output algorithm (HIO). We refer to it in what follows as ping-pong algorithm, which in software development commonly refers to algorithms that alternate between two strategies. The flow chart of this method is depicted in Appendix A.1.

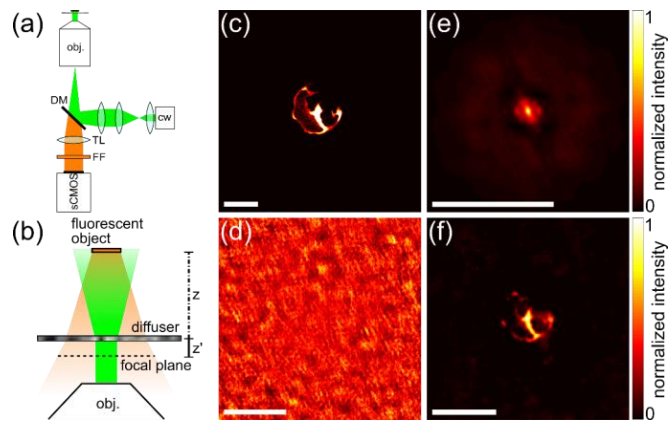

Fig. 1. (a) Regular inverted laser microscope in wide-field mode. cw:continuous wave light source, DM:dichroic mirror, TL:tube lens, FF:fuorescence filter. (b) Scheme of the excitation and fluorescence light diffusion with object-diffuser distance z and diffuser-focal plane distance z'. (c) Fluorescent object consisting of deposited 1 µm orange beads, imaged without diffuser (d) Normalized scattered light detected below the diffuser (distances z = 2 mm, z' = 600 µm). (e) Windowed autocorrelation of (d). (f) Reconstructed object using the ping-pong algorithm (see text). Scale bars are 200 µm in (c)-(e) and 50 µm in (f).

In all experiments the phase is restored from one single intensity image. Similarly to the HIO algorithm, the iterative algorithm is based on the constraint that the object intensity must be real and positive. However, adding an additional constraint, namely that the object's solution is searched in the real part of the complex solution (in the modified HIO algorithm) as well as in the absolute value of the complex solution (in the ER algorithm) in the object's domain. Furthermore, alternating between a modified version of the HIO and the ER prevents the algorithm to be stuck in local minima. This modified Fienup type algorithm takes four times less iterations than classical Fienup type algorithms and thus significantly improves the object reconstruction speed. More importantly it gives a more accurate and more reliable object's reconstruction (see Appendix A.2). Particularly in imaging scenarios with low signal-to-noise ratios, the ping-pong algorithm performs remarkably well (Appendix A.2), without the need to introduce additional image reconstruction priors [14].

Additionally, we report an alternative method to retrieve the Fourier domain modulus from the scattered light image (see Appendix A.3). In previous works, the Fourier transform of a windowed autocorrelation of the scattered light image was used to retrieve the object's Fourier domain modulus [9,12]. In contrast, we propose a direct Fourier transform of the scattered light image coupled with a smoothing filter that removes the speckle's information in the Fourier domain (See Appendix A.3). We noticed that this approach can present advantages under certain conditions, while giving similar object's reconstructions as obtained by the autocorrelation method. On the one hand, it can remove more efficiently the background that is residual from the speckle PSF, which can be advantageous when the object has to be reconstructed from small field of views with few speckle grains. On the other hand, insufficient sampling and subsequent smoothing can lead to a scrambling of frequency components that alter the object's shape in the reconstruction. Having both methods at hand thus increases the probability for a faithful reconstruction of the object that is hidden behind the scattering medium.

The results shown in Figs. 1(c)-(f) are obtained for a fluorescence micrometric structure placed at a distance z = 2 mm from the diffuser, and imaging the resulting speckle at z' = 600 µm from it. Interestingly, even complex features in the object can be reconstructed with high resolution. In what follows, we investigate the limits of these distances with respect to the quality of image reconstruction.

**Effect of the object-diffuser distance**

In the absence of a scattering medium, the fluorescent object can be directly imaged (as depicted in Fig. 2(a)) with high signal to noise conditions. In contrast, in the presence of a scattering medium, the excitation laser light and the emitted fluorescence light are affected equivalently by scattering, so that in both the excitation and the emission path the light flux decreases. Increasing distances of the object to the diffuser is a priori more favorable to ensure that the object size falls within the memory effect range of the scattering medium [13,17], however it results in less efficiently excited fluorophores and lower fluorescence light harvesting. Furthermore, in real bio-imaging experiments one does not have the luxury to choose the most convenient object-diffuser distance. It is important to notice that the scattering layer behaves like a magnifying lens depending on its distance from the object. This is due to the fact that, as a method based on the memory effect, it measures angles and not absolute positions. So an object closer to the scattering layer will cover a larger angle range than a farther one.

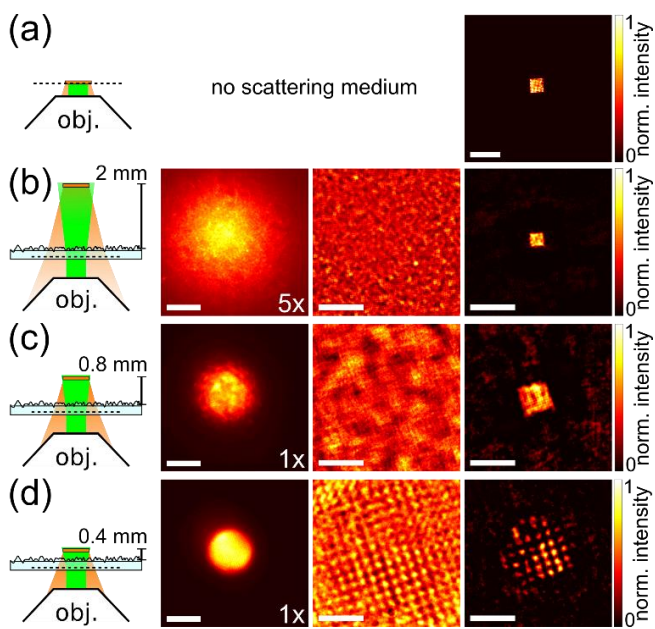

Fig. 2. Fluorescent quantum dot square pattern imaged with a 10x/0.25 objective. In the absence of a diffuser, scale bar 100 μm (a) or in the presence of a diffuser at different distances (b)-(d). The second column shows the raw scattered light image detected at 600 μm below the diffuser, the third column shows a cropped and normalized scattered light and the most right column shows the reconstructed object. Scale bars from second to fourth column are: 300 μm, 200 μm and 50 μm. Excitation power on the diffuser was 25 mW with integration times of 10 to 30 s.

At the same time the memory effect has only a finite range, and so imposes a finite angular field of view, so any object too close might be only partially visible. To evaluate this compromise we used a 40 μm size fluorescent object, made of quantum dots deposited in a square pattern, placed behind the diffuser at a varying distance from 2 mm down to 0.4 mm. The image plane was kept at a fixed position below the diffuser (Fig. 2). At object-diffuser distances smaller than 0.4 mm, the object could be imaged directly (when the focal plane of the objective was set to the object plane) with some minor distortions, due to the thin diffuser. This better fidelity to the object in the retrieved image is assigned to the so called shower curtain effect [18,19]. When the object was positioned at 2 mm (Fig. 2(b)) from the diffuser surface, the speckle recorded showed about 5 times lower intensity than for distances of 0.8 mm (Fig. 2(c)) and 0.4 mm (Fig. 2(d)). Increasing the input laser power or the camera acquisition times could be used to adapt the signal to noise ratio obtained, however an increase in input power was seen to generate visible auto-fluorescence from the scattering medium, which severely deteriorated the image contrast and made object reconstructions impossible.

Fig. 2(d) shows that, when the object is too close to the scattering layer and thus does not fit completely within the memory effect range, the edges of the object are not reconstructed, showing the presence of a finite field of view [12,20].

As the optical memory effect has a purely geometrical interpretation, for diffusive media its range depends only on the sample thickness. This is not true for optical diffusers, where light is on average only scattered once. We have estimated the angular memory effect of the ground glass diffuser used in this work by shifting the diffuser over a few hundred micrometers, fixing the object position (Appendix A.4). A minimum angle of 8.3° was obtained, meaning that a compromise of object size O and its distance from the diffuser z (here $O/z <\sim 0.15$), determines whether or not the object can be reconstructed, which is consistent with the distances observed in Fig. 2. Note that the obtained value is not far from that obtained in biological samples, which exhibit anisotropic scattering [12,20]. In fact the presence of a significant amount of forward scattering, common in many biological tissues, increases the memory effect range compared with the purely isotropic scattering case, which allows the observation of biological objects such as cells.

Another specificity of the epi-geometry studied here is that the object is itself illuminated by a speckle,

which might lead to artificial granularity in the reconstruction. At large distances z, the average excitation speckle grain size is $\delta_{exc} = z \cdot \lambda/D$, where $\lambda$ is the excitation wavelength and D is the diameter of the illuminated area on the diffuser. At distances z of a few hundreds of micrometers, the speckle is however not yet fully developed, which leads to a quite uniform illumination of the object. Breaking the coherence of the excitation light source would guarantee an even more uniform illumination of the object. Those results show that ultimately, the use of a thick scattering medium such as a biological tissue could be appropriate to image fluorescence labels buried inside at millimetric distances.

**Effect of the diffuser-objective distance**

In previously published experiments, the scattered light was detected at very large distances with a pinhole at the back of the scattering medium to ensure that the detected incoherent sum of speckles were correlated, placing the image plane in the far field to guarantee sufficient sampling and use the entire camera detection area [21,22]. Those geometries could only be used with high power laser sources, low numerical apertures and transmission objects.

Imaging micrometric scale fluorescent objects requires efficient light collection and magnifying lenses. Setting the focal plane of the objective close to the diffuser ensures an efficient fluorescence light collection (Fig. 3). The imaging magnification is however not anymore that of the objective, but must account for the scattering medium, which can be regarded as a lens [6] with a magnification of $M_{scatt} = z'/z$. Thus multiplication with the objective magnification $M_{obj}$ gives the total magnification of the system $M_{sys} = z'/z \cdot M_{obj}$ which determines the size of the image on the camera. Fig. 3 shows a fluorescent object of about 100 µm size, made of 1 µm fluorescent beads placed 2 mm away from the diffuser. The reconstructed object size grows as predicted with distance z', as a result of the magnification of the "scattering lens". Interestingly, small diffuser-camera distances of z'=150 µm already allow proper reconstruction, even though the regime of propagation is not yet a far field Fraunhofer regime. This situation also ensures high speckle contrast. Using longer distances z' can however be of interest for a better sampling of the image (see for instance Fig. 3(h) at z'=600 µm).

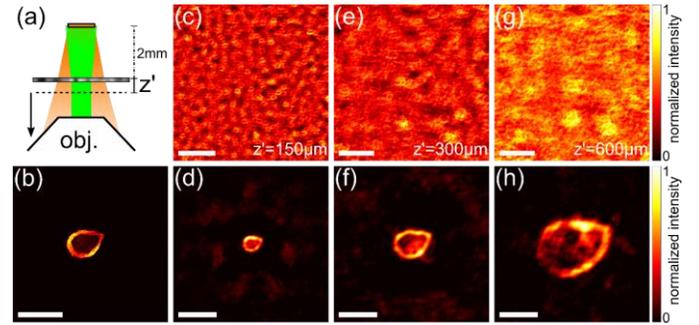

Fig. 3. Normalized scattered light image with the focal plane at three different distances from the diffuser, z' = 150, 300 and 600 µm. (a) Scheme of excitation and emission of the fluorescent object. The dashed line represents the focal plane of the objective. (b) Object imaged without diffuser. (c), (e) and (g) show the normalized scattered light. Scale bars: 100 µm. (d), (f) and (h) show the reconstruction of the object. Scale bars: 40 µm (in the plane of scattered light detection, using the scattering lens magnification Mscatt, see text). The sampling of the total object size along the horizontal direction is 34 (c), 69 (e) and 138 (g) pixels on the camera. Excitation power on the diffuser was 1 mW, integration time was 1 s.

For distances greater than 600 µm, the reconstruction failed, as can be seen for z'=600 µm where parts of the object are already missing. This is due to the fact that the further the focal plane is moved away from the diffuser, the more the scattered object replicas overlap within the formed speckle, leading to a loss of contrast. Moreover at large z' distances, longer integration times are required to compensate for the lower light flux.

At last, the combination of objective and scattering medium ultimately drives the resolution of the optical imaging method used here. It is primarily fixed by the speckle grain size in the object space, convolved by the objective optical resolution, which depends on its numerical aperture. Increasing the distance z' thus does not modify the global resolution but leads to better sampling.

**Effect of the objective magnification**

In an epi detection scheme, the objective always influences light excitation and light detection simultaneously. Therefore, the choice of magnification of the objective will influence the field of view (FOV) of the collected scattered light, but also the excitation FOV. Higher magnification objectives decrease the beam diameter of the illumination area on the scattering medium and therefore the excitation area of fluorescence objects placed close to the diffuser.

For the detection FOV, choosing a higher magnification leads to an extended use of the camera detection area and higher sampling of the imaged speckle, as can be seen in Figs. 2(c),(d) where the scattered light recorded with a 10x objective does not fill the camera chip.

**Effect of the emission spectral bandwidth**

In order for a light source to produce a contrasted speckle pattern, it needs to have both some degree of spatial and temporal coherence. As we are working with fluorescent objects spatial coherence is determined by the size and shape of the object itself. We explicitly use this fact to obtain Eqs. (1) and (2). Temporal coherence depends on the spectra bandwidth of the detected light. Wavelengths that are too far apart will generate completely different speckle patterns, which will average out reducing the contrast and eventually washing out any feature. The frequency range that will produce the same speckle pattern defines the spectral bandwidth of the medium [5,23]. It influences the measured speckle contrast [24] and is closely related to the long range correlations of the medium [25].

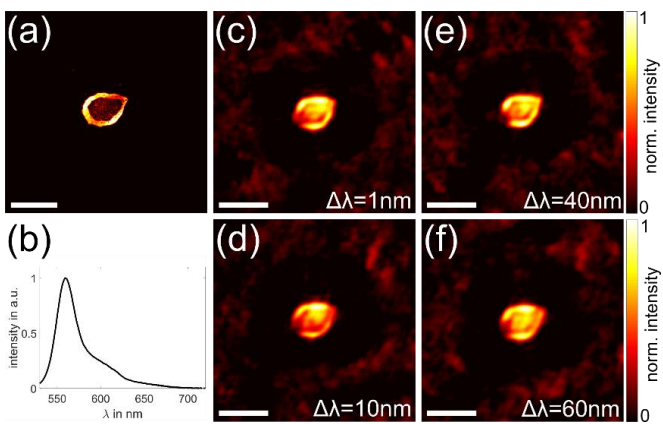

Fig. 4. Fluorescent object reconstructions with different emission filter bandwidths. (a) Object imaged without diffuser. (b) Emission spectrum of orange red beads. Reconstruction with bandpass filter 632-1 nm (c), 590-10 nm (d), 692-40 nm (e) and a 550 long pass filter (creating a 60 nm bandpass detection accounting for the dichroic mirror used) (f). Scale bars are 300 μm in (a) and 40 μm in (c) to (f). Excitation power on the diffuser was 1 mW with integration times from (c)-(d) of 30 s, 1.5 s, 5 s and 0.2 s respectively.

Fig. 4(c)-(f) shows the reconstruction of an object made of fluorescence beads (Fig. 4(a)) with large spectral emission (Fig. 4(b)), placed behind a diffuser, for different fluorescence emission bandwidths. The reconstructions were seen to be unaffected by an increase of the spectral bandwidths up to 40 nm, with slight degradation at 60 nm bandwidth (note that the background observed in the middle of the object is attributed to remaining background signal from the diffuser and not to the bandwidths used).

This result shows that large spectral bandwidths can be used in imaging behind a diffuser, ultimately enabling high photon flux to be recorded with optimal detection of the whole fluorescence spectrum of molecules. Note that spectral bandwidths of similar magnitudes have been observed in biological tissues [12]. This possibility is particularly important for future investigations using molecules of lower brightness or lower density than in the fluorescence beads demonstrated here. In the present work, auto-fluorescence background from the diffuser prevented to study such conditions, which can be reached by shifting the working wavelength more towards the red for instance.

**Conclusion**

We have shown how to reconstruct micrometer sized fluorescent objects through scattering media from a single intensity image of scattered light in a regular wide-field inverted microscope. First of all, collecting the scattered light close to the scattering medium, e.g. in the Fresnel zone, gives to the speckle images a high contrast that is appropriate for high quality reconstruction, in particular for the observation of complex objects. Interestingly in this regime, objects of various complexity could be satisfactorily reconstructed (Appendix A.5), including model objects similar to studied previously [12]. Second, both objective magnification and diffuser to objective distance govern the magnification of the reconstruction and efficient light harvesting. We have shown that in the context of biological samples investigations, taking advantage of the full emission bandwidth of fluorophores enhances the signal to noise ratio, which is crucial for the detection of micrometer sized objects and low power excitations. Together with a new combination of classical Fienup type phase retrieval algorithms that is computationally less heavy and turns out to be more robust in low SNR conditions, it is a step towards fluorescence bio-imaging through scattering samples. Future combination of the presented approaches with a guide star in the object space [8,26] or with a phase diversity approach [27] could also allow for deconvolution, depth-resolved fluorescence imaging.

**Funding**

S.B. was supported by a Leverhulme Trust Visiting Professorship. J.B. acknowledges support from the Leverhulme Trust's Philip Leverhulme Prize.

**Acknowledgments**

The direct laser written fluorescent 'F' pattern used in Fig. 11 was provided by Dr Patrick Salter (University of Oxford) and Dr Alex Corbett (University of Exeter).

## APPENDIX A

### A.1. Phase retrieval with the ping-pong algorithm

The Fourier domain phase $\varphi_G$ is wished to be retrieved correctly in order to reconstruct the object by simply inversely Fourier transform the found solution in the Fourier domain. To start the ping-pong algorithm (flow chart is depicted in Fig. 5), a random amplitude pattern $g_0$ is created. Both the hybrid input-output algorithm and the error reduction algorithm are Fienup type algorithms that rely on alternating projections between the object domain and the Fourier domain. Constraints in the Fourier domain and in the object domain lead the algorithm to converge to a solution. The modulus of the complex Fourier domain solution G is replaced by the modulus of the Fourier transform of the object $|H|=|F\{O\}|$. The knowledge of having a fluorescent amplitude object allows us to assume that the object must be real and positive.

Here, we use a slightly modified version of the hybrid input-output (HIO) algorithm in a sense that the pixels of g' = $F^{-1}\{|H|\cdot\exp(i\cdot\varphi_G)\}$ that violate the above stated assumptions are subtracted by a β-scaled version of the previous object domain solution g (in the regular HIO it is g - β·g'). The β value is incrementally decreased by 0.01 for each iteration k the HIO is restarted. It starts with 3 (fixed empirically) and the algorithm is stopped if β reaches 1. This leads to a total number of 201 HIO iterations. Each iteration of the HIO is followed by one iteration of the regular ER algorithm, where the pixels in the object domain are kept that have a positive absolute value and pixels that violate this condition are set to 0.

Consequently the modified HIO and the ER are computed alternately leading to a total of 402 iterations. The ping-pong algorithm thus takes around four times less iterations than comparable Fienup-type algorithms used in the field of speckle correlations [9,28–31].

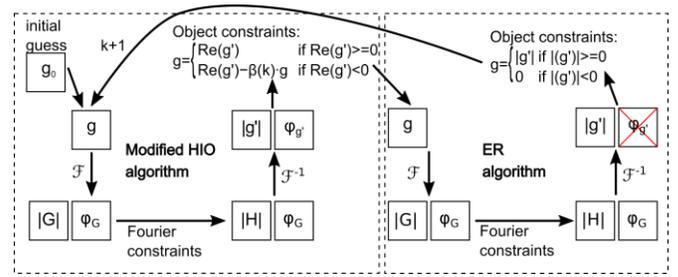

Fig. 5. Detailed flow chart of the ping-pong algorithm which consists of a modified hybrid input-output algorithm and the error reduction algorithm. The missing phase $\varphi_G$ is desired to be retrieved with |H| as the Fourier domain modulus as the only available information.

### A.2. Comparison of phase retrieval algorithms and image manipulation techniques in low signal-to-noise conditions

Fluorescence imaging can lead to loss of signal-to-noise ratio in conditions where the emitter is of low efficiency, or when using highly scattering media, which reduce both excitation and emission fluencies. For a comparison of the performance of algorithms under increasing noise conditions, it is necessary to introduce a metric that represents the quality of the reconstructed object. The error of the modulus in the Fourier domain is not a reliable metric because there are multiple solutions in the object domain that match the single intensity pattern in the Fourier domain. In the case of a priori knowledge of the object, the object itself can serve as a comparison. Therefore, we introduce the normalized cross correlation between the reconstructed object and the original object as a criterion to measure the quality of reconstruction.

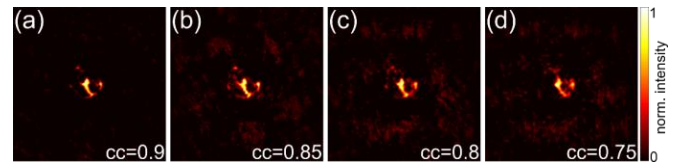

Fig. 6. Reconstructed objects from simulated scattered light images. The normalized cross correlation (cc) values are calculated with respect to the original object and serve as a metric to assess the quality of reconstruction.

Fig. 6 depicts four typical cases of reconstructions, of normalized cross correlation values (cc) that range from 0.9 to 0.75. Reconstructions with cross correlation values below 0.75 are considered to be solutions that show no resemblance with the original object. The aim of comparing existing Fienup type algorithms with the ping-pong algorithm proposed in this article is not only to show its performance but also to investigate how reliable the reconstructions are under different noise conditions, such that it might be possible in the future to image objects buried in scattering media without knowing their structure a priori.

Two Fienup type phase retrieval algorithms are compared - the hybrid input-output algorithm (HIO) available online [32] which we found to perform very well amongst others in this class of algorithms and the ping-pong (PP) algorithm presented in this article. The HIO is utilized with the β parameter decreasing from 2 to 0 in steps of 0.04 with each step iterated 30 times. After running the HIO, 30 iterations of the error reduction algorithm were performed. This leads to a total iteration number of 1560. Whereas the ping-pong algorithm was executed under the conditions stated in the section above with a total number of 402 iterations. Each of the phase retrieval algorithms was performed 20 times with different random initial guesses in the object domain. The mean value of cc for the 20 reconstructions is depicted in Fig. 7 as a blue circle (PP) and a red triangle (HIO). Error bars represent the mean deviation calculated separately for positive and negative deviations. Besides, the maximum value of cc reached for all reconstructions is marked with a square (PP) and a cross (HIO). Both algorithms were performed based on scattered light images with signal-to-noise (SNR) levels of 5, 8 and 15 dB, as well as for noise free scattered light images (∞dB) and experimental data.

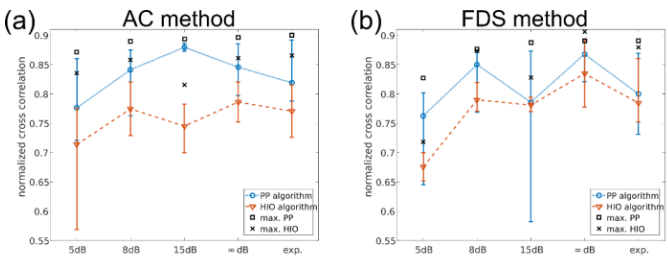

Fig. 7. Performance of a HIO algorithm with the ping-pong algorithm. The normalized cross correlation between the reconstruction and the original object is used a reconstruction quality metric. The mean value over 20 reconstructions is plotted as a blue circle (PP) and a red triangle (HIO) for signal-to-noise values of the scattered light image of 5, 8 and 15 dB, as well as for noise free (∞dB) and experimental data. Error bars indicate the mean deviation in both positive and negative direction. The square and cross sign indicate the maximum normalized cross correlation value reached for each of the reconstruction conditions. Additionally, the autocorrelation (AC) and Fourier domain smoothing (FDS) image treatment techniques are compared.

The noisy images were created by using a noise model specially developed for CMOS cameras [33] that accounts for multiple sources of noise (i.e. photon shot noise, photo response non-uniformity, dark current shot noise, flicker noise, quantization noise etc.) with parameters that resemble the camera used in the articles' experiments. In high noise conditions, a low-pass filter was applied in the Fourier domain to ensure the suppression of high frequency components that can be fully attributed to noise. Additionally, the quality of reconstruction is compared between two image manipulation methods – the autocorrelation technique (AC) and the Fourier domain smoothing technique (FDS, explained in the next section) – that are given in the chart of Fig. 7(a) and (b) respectively.

Independent of how the Fourier domain modulus from the scattered light image was retrieved and if noise is present in the images or not, the ping-pong algorithm outperforms the classical HIO algorithm, in addition of taking four times less iterations. The difference is most significant at very low SNR levels of 5 and 8 dB where the mean cc-value of the PP algorithm reconstructions are between 0.06 and 0.1 higher as compared to the HIO reconstructions. Thus, even at low signal-to-noise levels of 5 dB, reconstructions with the PP algorithm allow for a relatively good object reconstruction (depicted in Fig. 8(a)). In low SNR cases, high frequency components are hidden in the noise and therefore the reconstruction will appear as a low pass filtered version of the object. This can be observed progressively when looking at the reconstructions from the noise-free image down to the 5 dB images (Fig. 8(a)-(d)).

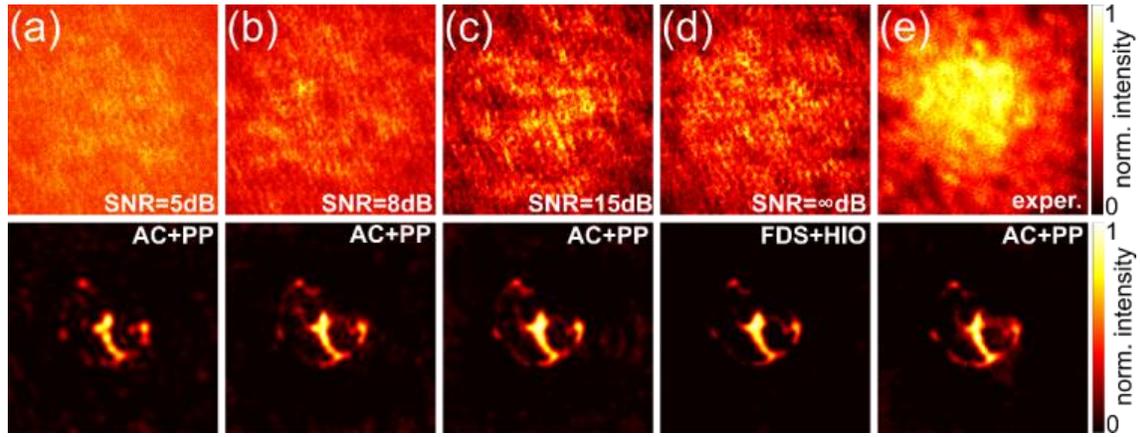

Fig. 8. Best reconstruction results for varying signal-to-noise levels on the scattered light images (left row) (a)-(d) and experimental data (e). The best combination of image treatment (AC or FDS) and phase retrieval algorithm (PP or HIO) is indicated in the top right corner of the Fourier domain modulus (left row) and the reconstruction (right row) respectively.

Here only the image retrieved with maximum cc value is depicted. In all cases, except one, a combination of AC method and PP algorithm gave best reconstruction results even though the combination of FDS method and PP algorithm reached similar values as compared to the reconstructions depicted in Fig. 8. In the case of the noise-free image (Fig. 8(d)), the combination of Fourier domain smoothing (FDS) and HIO algorithm gave the best result, showing the better robustness of the PP algorithm with respect to noise.

Note that in average, the comparison of the AC and FDS approaches does not give clear trends when using the PP algorithm. When using HIO however, FDS shows higher qualities of reconstructions above 5dB SNRs. In a noise-free situation, the FDS method performs better independent of which algorithm was used (PP or HIO).

Both the AC and the FDS methods are very sensitive to careful sampling, windowing, smoothing and low-pass filtering. Therefore, the advantage of the FDS method is to provide a complementary approach to retrieve high quality reconstructions in case of the failure of the AC method or vice versa.

**A.3. Image treatment in the Fourier domain**

The most common method to remove the speckle's contribution from the detected scattered light image is to apply an autocorrelation and use the Wiener-Khinchin theorem to reconstruct the object. However, windowing the autocorrelation leaves spurious spatial frequencies of the speckle's info and sometimes introduces artefacts in the reconstruction.

A complementary approach to remove the speckle's contribution will be referred to as the Fourier domain smoothing (FDS) method that is described in the following.

In incoherent illumination it is known that the PSF of a scattering medium is an intensity speckle (Fig. 9(b)). Assuming we have an object O (Fig. 9(a)) behind the scattering medium, the image I that we would hypothetically measure at a distance from the scattering medium is the convolution of the object O with the speckle PSF S: $I = S \otimes O$ (Fig. 9(c)). In the Fourier domain the convolution operation turns into a multiplication and by directly Fourier transforming the scattered light image of Fig. 9(c) it can be seen that the Fourier amplitude of the object is maintained but affected by the speckle's Fourier transform. The latter is depicted in Fig. 9(e). At this stage in the direct FT image, there is no obvious retrieval of the object, due to the speckle information contained. However, if we smoothen the Fourier domain modulus of a speckle (Fig. 9(e)) with an average filter (or Gaussian filter), what is remaining is a uniform intensity envelope with no information left on the speckle (Fig. 9Fig. (g)). Sufficient sampling during the FFT performance guarantees indeed that the frequency mixing is limited and therefore the object's shape can be maintained. This can be seen by comparing the true FT of the object in Fig. 9(d) with the smoothened FT of the object taken from the scattered light image in Fig. 9(h).h

To validate this approach, the object is reconstructed with the correct object phase in the Fourier domain taken from the FT of the object and applied to the smoothened FT of the scattered light image from Fig. 9(h). The result is shown in Fig. 9(i) – except an expected loss of resolution, the object is satisfactorily reconstructed. Eventually, by taking the retrieved phase from the ping-pong (or alternatively HIO) algorithm, the object can be reconstructed very accurately (Fig. 9(j)). This confirms that the smoothing in the Fourier domain efficiently removes the PSF's speckle noise and in the same time does not disturb the object's Fourier domain modulus so that a reliable object reconstruction can be achieved. For comparison of the simulated images, the smoothened Fourier domain modulus of experimental data is depicted in Fig. 9(k) and the deducted reconstruction in Fig. 9(l). Therefore, experimental results confirm the validity of the FDS method as an appropriate choice for scattered light image treatment and its object's reconstruction. It can in particular allow a higher degree of control over possible parasite sources of signal such as scattering background for instance, since spatial frequencies are more distinctly manipulated.

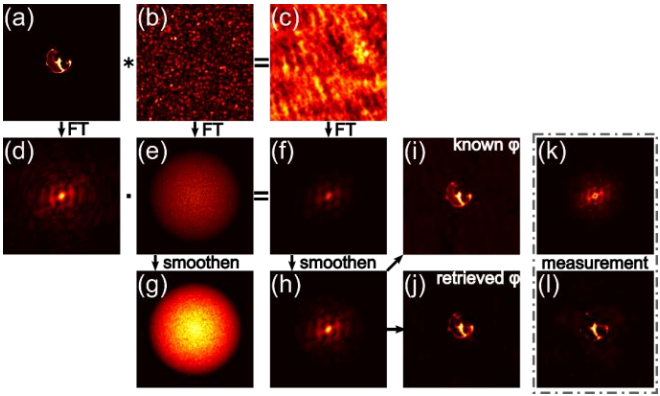

Fig. 9. (a) object, (b) generated speckle, (c) convolution of object with speckle, (d) FT of object, (e) FT of speckle, (f) FT(I)=FT(S)*FT(O), (g) smoothened speckle FT, (h) smoothened FT(I), (i) reconstruction of (h) with phase retrieval algorithm, (j) IFT of (h) with known phase of object, (k) smoothened FT(I) from experimental data, (l) reconstruction from (k).

### A.4. Memory effect of a ground glass diffuser

To estimate the angular memory effect of the diffuser, a sample made of fluorescence beads into a structure of 200x300 μm was placed 2 mm behind the diffuser and the scattered light was collected 150 μm below the diffuser. The diffuser was translated within the field of view of the objective and camera (665 μm) to ten different positions from its initial position while the fluorescent beads object was kept fixed in position, which to a certain extent is similar to translating the object with respect to the diffuser. Fig. 10 shows the normalized cross correlation of the obtained speckle image, calculated for all possible translations which are turned into an equivalent tilt angle that is seen from the diffuser. This angle is deduced from the translation distance d of the diffuser, calculated by arctan(d/z), where z is the object between object and diffuser. In this geometry, scattered light images are correlated until the diffuser is translated by 300 μm, leading to an angular memory effect of δ = arctan(300 μm / 2000 μm) = 8.53°.

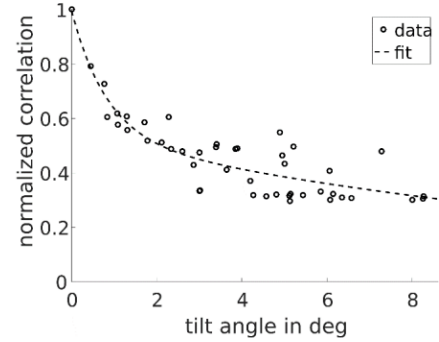

Fig. 10. Correlation of scattered light images recorded with the diffuser translated. The tilt angle is calculated from the translation distance d of the diffuser by arctan(d/z), where z is the distance between object and diffuser.

### A.5. Object's complexity

A detailed discussion about the performance of reconstruction algorithms with respect to the object's complexity can be found in Katz et al. [28]. As a complementary point, we observed scattered light field imaged so close to the diffuser surface that the far field is not reached yet. In this Fresnel region, the speckle grain size is comparable to the object's dimensions which gives the highest contrast in the speckle used for the reconstruction of the object. Therefore, more complex objects can be reconstructed because the high contrast permits the distinction of structural details that are within the dynamic range of the camera. Fig. 11(a) depicts a low complexity letter 'F' pattern, fabricated by direct laser writing in a clear polymer, similar to the process described in [34], and an object formed by orange beads in (Fig. 11(b)) that is more complex. The center row depicts the scattered light image from which the object's FT modulus was retrieved and the right row shows the reconstructed object with the phase obtained by the ping-pong algorithm. Importantly, the algorithm performs as good in complex and simpler object shapes.

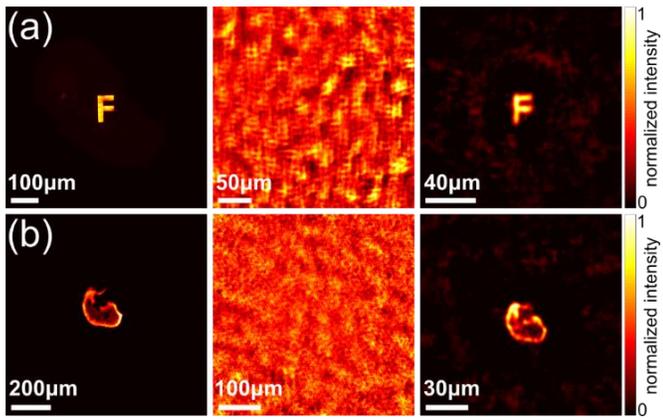

Fig. 11. A simple (a) and a more complex object (b), its scattered light after the diffuser (center row) and the reconstruction (right row). (a) 'F' pattern fabricated by direct laser writing process with organic dyes fluorescent structures in a clear polymer. (b) Drop casted orange beads.